\documentclass[prl,reprint,amssymb,superscriptaddress]{revtex4-2}

\usepackage{xcolor}

\usepackage{bm}

\usepackage{amsmath}  
\usepackage{amsfonts}
\usepackage{graphicx}
\usepackage{hyperref}
\hypersetup{
	colorlinks=true,
	citecolor=blue,
	linkcolor=red,
	urlcolor = blue
}

\begin{document}
	
	\title{Reconciling Kubo and Keldysh Approaches to Fermi-Sea-Dependent Nonequilibrium Observables: Application to Spin Hall Current and Spin-Orbit Torque in Spintronics}
	
	\author{Sim\~{a}o M. Jo\~{a}o}
	\affiliation{Department of Materials, Imperial College London, London SW7 2AZ, UK}
	
	\author{Marko ~D.~Petrovi\'c}
	%\altaffiliation{Present address: Department of Physics, Georgetown University, Washington, DC 20057, USA}
	\affiliation{Department of Physics and Astronomy, University of Delaware, Newark, DE 19716, USA}

	\author{J. M. Viana Parente Lopes}
	\affiliation{Centro de Física das Universidades do Minho e Porto and Departamento de F\'{i}sica e Astronomia,
		Faculdade de Ci\^{e}ncias, Universidade do Porto, 4169-007 Porto, Portugal}
	
	\author{Aires Ferreira}
	\email{aires.ferreira@york.ac.uk}
	\affiliation{Department of Physics, University of York, YO10 5DD, York, UK}

	\author{Branislav K. Nikoli\'c}
	\email{bnikolic@udel.edu}
	\affiliation{Department of Physics and Astronomy, University of Delaware, Newark, DE 19716, USA}

	\begin{abstract}
		Quantum transport studies of spin-dependent  phenomena in solids  commonly employ the
		Kubo or Keldysh formulas for the  nonequilibrium density operator  in the steady-state linear-response regime. Its trace with  operators of interest, such as the spin density, spin current density, etc., gives expectation values of experimentally accessible observables. For   local quantities, these formulas  require summing over the manifolds of {\em both} Fermi-surface and Fermi-sea  states. However, debates have been raging in the literature about the vastly different physics the two formulations can  apparently  produce, even when applied to the same system. Here, we revisit this problem  using  infinite-size  graphene with proximity-induced spin-orbit  and magnetic exchange effects as a testbed. By splitting this system into semi-infinite  leads and central active region, in the spirit of Landauer formulation of quantum transport, we prove the {\em numerically exact equivalence} of the Kubo and Keldysh approaches via the computation of spin Hall current density and spin-orbit torque in  both clean and disordered limits. The key to reconciling the two approaches are the numerical frameworks we develop for: ({\em i}) evaluation of Kubo(-Bastin) formula for a system attached to semi-infinite leads, which ensures continuous energy spectrum and evades the need for commonly used  phenomenological broadening introducing ambiguity; 
		and ({\em ii}) proper evaluation of Fermi-sea  term in the Keldysh approach, which {\em must} include the voltage drop across the central active region even if it is disorder free. 
	\end{abstract}
	
	\maketitle
	
	The density (or state or statistical) operator~\cite{Ballentine2014} is the central concept of quantum statistical mechanics. The operator and its representation---the density matrix---in some basis of the Hilbert space is {\em sine qua non} to describe  equilibrium quantum systems at finite temperature~\cite{Kardar2007}, as well as  
	out-of-equilibrium systems driven by steady or time-dependent external fields. The density operator also plays a crucial role in describing the transition between nonequilibrium and equilibrium states~\cite{Nathan2024}, as well as in zero-temperature quantum mechanics~\cite{Weinberg2014} and quantum information science where it describes decoherence (i.e., the decay  of the off-diagonal elements of the density matrix in some preferred  basis~\cite{Joos2003}), and eventually quantum-to-classical transitions \cite{Joos2003,GarciaGaitan2024a}. From the knowledge of the density matrix $\hat{\rho}$, the expectation value of a physical observable represented by a  Hermitian operator $\hat{O}$ is obtained from $O=\mathrm{Tr}\, [\hat{\rho} \,\hat{O}]$. The density matrix in equilibrium is  unambiguously specified  by the Boltzmann and Gibbs prescription. A textbook example~\cite{Kardar2007} is the grand canonical ensemble \mbox{$\hat{\rho}_\mathrm{eq} = \sum_n f(E_n)|E_n\rangle \langle E_n|$} for noninteracting fermions with single-particle energy levels $E_n$, occupied according to the Fermi-Dirac distribution function $f(E)$, and with the corresponding eigenstates $|E_n\rangle$ of a fermionic Hamiltonian. By contrast, there is no unique procedure to obtain the density matrix of steady-state~\cite{Dutt2011,Dhar2012} 
	or time-dependent~\cite{Gaury2014,Popescu2016,Bajpai2020,Ridley2022}   nonequilibrium systems and for arbitrary strength of the driving field~\cite{Dutt2011}, while including their many-body interactions~\cite{Ridley2022}.
	
\begin{figure}
		\includegraphics[scale=1.0]{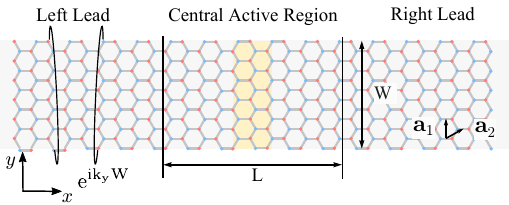}
		\caption{Doubly-proximitized~\cite{Zollner2020} infinite   graphene sheet viewed as a two-terminal Landauer setup~\cite{Imry1999,Nazarov2009,Baranger1989,Caroli1971,Fisher1981} for computational quantum transport~\cite{Waintal2024} with its CA region being an armchair nanoribbon (of finite length \mbox{$L = 40$ \AA} and width \mbox{$W = 15$ \AA}) attached to semi-infinite nanoribbons of the same kind. The nanoribbon  is  periodically repeated  along the $y$-axis to reproduce bulk behavior of an infinite sheet~\cite{Liu2012d}. We employ this setup  in the calculation of SH current density [Fig.~\ref{fig:fig2}] and SO torque [Fig.~\ref{fig:fig3}], within the shaded-in-yellow middle strip,  via both Kubo Eq.~\eqref{eq:rhokubo} and
			Keldysh Eq.~\eqref{eq:rhokeldyshlr}.}
		\label{fig:fig1}
	\end{figure}

	Nonetheless, in problems of noninteracting electrons driven by weak external fields, linear-response theory of Kubo~\cite{Kubo1957} or Keldysh Green's functions (GFs)~\cite{Keldysh1965,Stefanucci2013} makes it possible to construct a {\em universal} expression for $\hat{\rho}^\mathrm{neq}$ that is expressed in terms of the retarded GF involving Hamiltonian of the system in {\em equilibrium}, $\hat H$. For example, in the case of Kubo formula~\cite{Kubo1957}, and using Bastin {\em et. al.}~\cite{Bastin1971} formulation in terms of GFs, the steady-state $\hat{\rho}^\mathrm{neq}$ (at zero temperature for simplicity) is given by 
	\begin{subequations}\label{eq:rhokubo}
		\begin{eqnarray}
			\hat{\rho}^\mathrm{neq}_{\rm Kubo} & = & {\hat{{\rho}}}^{\rm surf}_{\rm Kubo} + {\hat{{\rho}}}^{\rm sea}_{\rm Kubo}, \\
			{\hat{{\rho}}}^{\rm surf}_{\rm Kubo} & = & 
			\frac{4eE_x}{h} {\rm Re}\left[
			\mathrm{Im}\, \hat{G} \,
			\hat{v}_{\rm x}\,
			\mathrm{Im}\, \hat{G}
			\right], \label{eq:rhokubosurf} \\
			\hat{\rho}^{\rm sea}_{\rm Kubo} & = &
			\frac{2eE_x}{h} \int\!\!
			dE f(E)
			{\rm Re}\left[
			(\hat{G}^{r} - \hat{G}^{a})
			\hat{{v}}_{\rm x}\,
			\partial_{E} \mathrm{Re}\, \hat{G}
			\right], \label{eq:rhokubosea}
		\end{eqnarray}
	\end{subequations}
	where $\partial_E \equiv \partial/\partial E$ and $E_x$ is the strength of the small electric field (assumed to point along the  $\hat x$-axis). Note that several decompositions~\footnote[1]{See Supplemental Material (SM) at \url{https://wiki.physics.udel.edu/qttg/Publications}, which includes Ref.~\cite{Mousavi2012}, for: (i) summary of different commonly employed decompositions of Kubo Eq.~\eqref{eq:rhokubo} and Keldysh Eq.~\eqref{eq:rhokeldyshlr}, as well as additional details on their numerical evaluation; (ii) one additional figure recalling a well-established~\cite{Baranger1989,Nikolic2001b} equivalence of Kubo and Keldysh formulas for conductance (i.e., longitudinal charge transport) calculations using an example of plain [i.e., without SOC and magnetism terms in Eq.~\eqref {eq:hamiltonian}] graphene in Fig.~\ref{fig:fig1}; and (iii) for the same plain graphene, we also show that its density of states computed using the setup of Fig.~\ref{fig:fig1} is identical to standard result~\cite{CastroNeto2009a} for infinite graphene.} into  ${\hat{{\rho}}}^{\rm sea}_{\rm Kubo}$ and ${\hat{{\rho}}}^{\rm surf}_{\rm Kubo}$, governed by the Fermi-sea and Fermi-surface states, respectively, have been used  historically~\cite{Smrcka1977,Streda1982,Crepieux2001,Bonbien2020,Ado2024}. Here we employ the specific ``symmetrized  decomposition'' of Ref.~\cite{Bonbien2020}  ensuring no overlap between the two terms, 
	while also using ${\hat{{\rho}}}^{\rm surf}_{\rm Kubo}$  as the Kubo-Greenwood form~\cite{Baranger1989,Greenwood1958} which is advantageous for comparison with Keldysh Eq.~\eqref{eq:rhokeldyshlrsurf}.  The retarded ($r$) GF standardly used in Eq.~\eqref{eq:rhokubo} is given by 
	\begin{equation}\label{eq:retardedgf}
		{\hat{G}}^r = \left[E - \hat{H} + i\eta \right]^{-1},
	\end{equation}
	where  $\hat{G}^a=(\hat{G}^r)^\dagger$ is the advanced GF; $\mathrm{Im}\, \hat{G} = (\hat{G}^r - \hat{G}^a)/2i$; $\mathrm{Re}\,\hat{G}=(\hat{G}^{r}+\hat{G}^{a})/2$,  $\hat{\mathbf{v}} =   [\hat{\mathbf{r}}, \hat{H}]/i\hbar$ is the velocity operator and $\hat{\mathbf{r}}$ is the position operator.  In most practical applications employing atomistic models~\cite{Joao2020,Fan2021}, $\hat H$ is either a generic symmetry-allowed~\cite{Ghosh2018,VanTuan2016,Garcia2017,MedinaDuenas2024} or first-principles-derived~\cite{Freimuth2014,Mahfouzi2018,Mahfouzi2020,Santos2018}  tight-binding  Hamiltonian   on a {\em finite} lattice with {\em periodic} boundary conditions. The parameter $\eta$ (formally an infinitesimal) is required  to define $\hat{G}^{r,a}$ by avoiding poles on the real axis~\cite{Economou2006}, with the limit $\eta \rightarrow 0$ taken explicitly in analytical approaches~\cite{Dimitrova2005,Ado2015,Milletari2016,Milletari2017,Perkins_2024}. However, in numerics $\eta$ must remain  nonzero due to the discreteness of energy levels ~\cite{Imry_2002,Nomura2007,MedinaDuenas2024},  and it can be reduced only at the computational expense of handling larger systems or averaging over random twisted boundary conditions \cite{Pires2022}. Physically, a finite $\eta$ was initially interpreted as mimicking the effect of uncorrelated inelastic scattering processes, thereby defining a phenomenological dephasing timescale $\tau_\phi =\hbar/\eta$~\cite{Thouless1981}. Later on, $\eta$ has  often been  interpreted~\cite{Freimuth2014,Bonbien2020} as a homogeneous broadening due to scattering from short-range impurities, but disorder should be explicitly introduced as a term in $\hat{H}$ [such as through $\varepsilon_i$ in  Eq.~\eqref{eq:hamiltonian}] in order to capture the same key concepts, disorder self-energy  and vertex corrections, used in  analytical calculations~\cite{Dimitrova2005,Ado2015,Milletari2016,Milletari2017,Joao2022,Perkins_2024}. Recently, inspired by the success of  approximation theory \cite{Boyd2000,Weisse_06}, a more practical and accurate interpretation has arisen, where $\eta$ is seen as an energy resolution that determines how close finite-size spectral simulations are to describe genuine thermodynamic behavior~\cite{Ferreira2015,Joao2020,Brito_24,Bulanchuk2021}. 
	
	The Keldysh formalism~\cite{Keldysh1965,Stefanucci2013} is applicable beyond the linear-response regime by employing additional GFs. Its fundamental quantities are the retarded $\hat{G}^r(t,t')$ and the lesser $\hat{G}^<(t,t')$ GFs describing  the density of available quantum states and how electrons occupy those states, respectively. The diagonal-in-time  component of the latter yields the  time-dependent nonequilibrium density matrix according to $\hat{\rho}^\mathrm{neq}_\mathrm{Keldysh}(t) = \frac{\hbar}{i} \hat{G}^<(t,t)$ \cite{Gaury2014}. In steady-state out-of-equilibrium scenarios, all quantities depend only on the difference $t-t^\prime$, so that after a Fourier transformation in energy domain $E$, $\hat{G}^<(E)$ yields the Keldysh formula for the nonequilibrium density matrix, \mbox{$\hat{\rho}^\mathrm{neq}_\mathrm{Keldysh} = \frac{1}{2\pi i} \int dE \, \hat{G}^<(E)$}. This formula is widely used in computational quantum transport~\cite{Waintal2024} studies  of two-~\cite{Brandbyge2002,Nikolic2005d,Reynoso2006,Haney2007,Areshkin2009} or multi-terminal~\cite{Nikolic2006,Gulbrandsen2020} Landauer setups~\cite{Imry1999,Imry_2002,Nazarov2009,Caroli1971,Baranger1989,Fisher1981,Waintal2024}, where a finite-length central active (CA) region, as in Fig.~\ref{fig:fig1}, is  coupled to  macroscopic Fermi liquid reservoirs via ideal  semi-infinite  leads. For instance, in the two-terminal geometry in which the left ($L$) and right ($R$) lead terminate into the corresponding  reservoirs characterized by the Fermi-Dirac functions, $f_{L,R}=f(E-\mu_{L,R})$, and in the presence of a bias voltage $eV_b=\mu_L - \mu_R$ [here, $\mu_{L,R}$ denotes the electrochemical potential~\cite{Payne1989} of $L$ or $R$ lead],  
	the lesser GF of the CA region was expressed by Caroli {\em et al.}~\cite{Caroli1971} as \mbox{$\hat{G}^<(E)  =  i \hat{G}^r(E)\left[f_L(E)\hat{\Gamma}_L(E) + f_R(E)\hat{\Gamma}_R(E) \right]\hat{G}^a(E)$}. The simplicity of this expression  stems from  the assumption of  many-body interactions being absent both in the CA region ~\cite{Meir1992,Ness2010,Mahfouzi2014} and in the leads~\cite{Ness2011}. 
	Here the retarded GF 
	\begin{equation}\label{eq:retardedgfleads}
		\hat{G}^r(E)  =   \left[E  - {\hat{H}} - \hat{\Sigma}_L^r(E) -\hat{\Sigma}_R^r(E) \right]^{-1},
	\end{equation}
	{\em differs} from Eq.~\eqref{eq:retardedgf} used in typical numerical Kubo calculations on systems with periodic boundary conditions~\cite{Freimuth2014,Fan2021,Mahfouzi2018,Mahfouzi2020,Ghosh2018,Gulbrandsen2020,MedinaDuenas2024} as it incorporates the leads  through their self-energies~\cite{Velev2004,Rungger2008,Waintal2024}  $\hat{\Sigma}_{L,R}^r(E)$. They define  the level broadening operators, \mbox{$\hat{\Gamma}_{L,R}(E)= i[\hat{\Sigma}_{L,R}^r(E)-\hat{\Sigma}_{L,R}^a(E)]$}, quantifying the electron escape rate into the leads. 

	The Keldysh formula  is valid in the nonlinear regime (i.e., for large  $V_b$)~\cite{Brandbyge2002,Areshkin2009,Ellis2017}. Thus, to compare it with the Kubo Eq.~\eqref{eq:rhokubo}, we assume small 
	$eV_b \ll E_F$, where $E_F$ is the Fermi energy, leading to $f_L - f_R \approx eV_b \, \delta(E-E_F)$  at zero temperature. In such linear-response limit, $\hat{\rho}^\mathrm{neq}_{\rm Keldysh}$ can be decomposed~\cite{Brandbyge2002,Mahfouzi2013} as (many other decompositions~\footnotemark[1] are possible~\cite{Mahfouzi2016,Nikolic2018})
	\begin{subequations}\label{eq:rhokeldyshlr}
		\begin{eqnarray}
			\hat{\rho}^\mathrm{neq}_{\rm Keldysh} & = & {\hat{{\rho}}}^{\rm surf}_{\rm Keldysh} + {\hat{{\rho}}}^{\rm sea}_{\rm Keldysh}, \\
			{\hat{{\rho}}}^{\rm surf}_{\rm Keldysh} & = & 
			\frac{eV_b}{2\pi} \hat{G}^r \hat{\Gamma}_L \hat{G}^a, \label{eq:rhokeldyshlrsurf} \\
			\hat{\rho}^\mathrm{sea}_\mathrm{Keldysh} & = & 
			- \frac{1}{\pi} \int\limits_{-\infty}^{+\infty}\!\! dE f_R(E)  \mathrm{Im}\, \hat{G}_\mathrm{V_b}(E) \nonumber \\ 
			\mbox{}	&& + \frac{1}{\pi} \int\limits_{-\infty}^{\infty} dE f(E) \mathrm{Im}\, \hat{G}(E).
			\label{eq:rhokeldyshlrsea}
		\end{eqnarray}
	\end{subequations}
	In Eq.~\eqref{eq:rhokeldyshlrsea} we explicitly subtract  the grand canonical density matrix in equilibrium (which is built in into Kubo's  derivation~\cite{Kubo1957}), $\hat{\rho}_\mathrm{eq} = -\frac{1}{\pi} \int\limits_{-\infty}^{\infty} dE f(E) \mathrm{Im}\, \hat{G}(E)$ expressed in terms of GFs, as  commonly done~\cite{Ellis2017,Theodonis2006,Nikolic2006} to remove expectation values that can be nonzero in equilibrium but are experimentally not observed, such as a field-like~\cite{Ralph2008,Belashchenko2019}  component of spin  torque~\cite{Ellis2017,Theodonis2006}, circulating spin currents~\cite{Nikolic2006,Rashba2003}, and circulating charge current~\cite{Baranger1989} (in the presence of external magnetic field).
	
	It may seem at first that the Kubo Eq.~\eqref{eq:rhokubo} and Keldysh Eq.~\eqref{eq:rhokeldyshlr} are of quite different form  and describing  quite different scenarios, that is, bulk transport in the former and two-terminal setups in the latter. Nonetheless, the two formulations are general and should lead to the same  physical conclusions. Indeed,  their equivalence~\footnotemark[1]  is well-established~\cite{Fisher1981,Baranger1989} for longitudinal charge transport observables requiring only $\hat{\rho}_\mathrm{Kubo}^\mathrm{surf}$ or $\hat{\rho}_\mathrm{Keldysh}^\mathrm{surf}$. 
	However, notably in the context of spin transport, the use of the two formulations has generated widespread confusion and even highly {\em divergent conclusions} about the same phenomenon. For example, Ref.~\cite{Gulbrandsen2020} compared the spin Hall (SH)  current obtained from Keldysh calculations in 4-terminal geometry~\cite{Nikolic2006} to that of Kubo calculations on periodic lattices, concluding that the Kubo formula is insufficient. Conversely, in spin-orbit (SO) torque calculations, the Keldysh approach  apparently predicts~\cite{Kalitsov2017,Zollner2020} only the field-like (i.e., odd in magnetization~\cite{Belashchenko2019,Dolui2020}) component of SO torque $\mathbf{T}_\mathrm{Keldysh} \equiv \mathbf{T}^o$ in the clean limit, while Kubo formula yields~\cite{Kurebayashi2014,Lee2015,Li2015,Sousa2020,Veneri_22} a nonzero value of both $\mathbf{T}^o$ and  damping-like (i.e., even in magnetization~\cite{Belashchenko2019,Dolui2020}) SO torque $\mathbf{T}^e$. The even SO torque from the Kubo formula has a well-studied ``intrinsic'' contribution that is governed~\cite{Kurebayashi2014,Lee2015,Li2015,Freimuth2014} by the Berry curvature of occupied Fermi-sea states and is largely insensitive to a phenomenological $\eta$ or  even to real-space disorder~\cite{MedinaDuenas2024}.  Thus, $\mathbf{T}_\mathrm{Keldysh} \neq \mathbf{T}_\mathrm{Kubo}$, even when both calculations are performed on an identical system, such as the paradigmatic Rasbha SO- and exchange-coupled two-dimensional (2D) electron  gas~\cite{Kalitsov2017,Li2015,Lee2015}. This in turn has led to the opposite perception, that is, that the Keldysh formula is insufficient. 
	However,  $\eta \rightarrow 0$ leads to $\mathbf{T}^o \rightarrow \infty$ [Fig.~\ref{fig:fig3}(f)] in Kubo calculations of clean systems, which led to concerns~\cite{Kalitsov2017} about the use of Kubo Eq.~\eqref{eq:rhokubo} in the absence of extrinsic scattering mechanisms. Additionally, calculations via  Kubo Eq.~\eqref{eq:rhokubo} with  first-principles Hamiltonian $\hat{H}$ of ferromagnet/heavy-metal bilayers, such as Co/Pt,  plugged into  GF in Eq.~\eqref{eq:retardedgf} {\em do not}~\cite{Freimuth2014,Mahfouzi2018} find strongly anisotropic angular features of SO torque, thereby contradicting  calculations~\cite{Belashchenko2019} using Keldysh Eq.~\eqref{eq:rhokeldyshlr} with GF in Eq.~\eqref{eq:retardedgf} or experiments~\cite{Garello2013} on the same Co/Pt system.

	\begin{figure}
		\includegraphics[width=\linewidth]{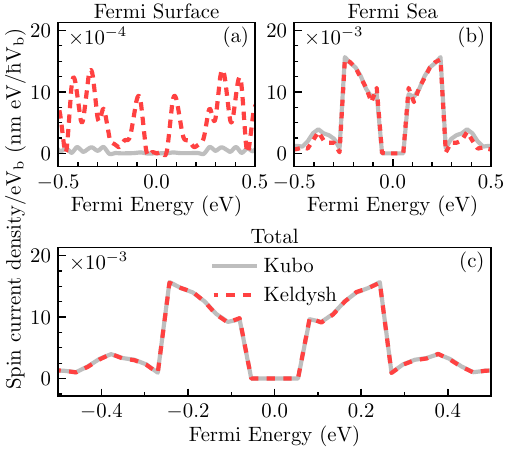}
		\vspace{-0.3in}	\caption{Spin Hall current density---obtained by tracing its operator~\cite{Wang2016a} with (a) Fermi-surface, (b) Fermi-sea, and (c) total density matrices---in Kubo [Eq.~\eqref{eq:rhokubo}] vs. Keldysh [Eq.~\eqref{eq:rhokeldyshlr}] approaches employing the same retarded GF [Eq.~\eqref{eq:retardedgfleads}] of doubly-proximitized graphene.  The parameters in Eq.~\eqref{eq:hamiltonian} are set as   $\lambda_\mathrm{RSO}=J_{\text{sd}}=0.1$ eV and the disorder strength is \mbox{$D=0.3$ eV}. The spin current density is averaged over $200$ disorder configurations. Convergence with respect to $k_y$-point sampling was also established.}
		\label{fig:fig2}
	\end{figure}

	In this Letter, we demonstrate the {\em numerically exact equivalence} between the Kubo  and Keldysh approaches by focusing on two paradigmatic examples of coupled spin-charge transport phenomena in spintronics, i.e., the SH effect and the SO torque. We provide a constructive proof of their equivalence by developing numerical frameworks which
	({\em i}) apply  the Kubo(-Bastin)  density matrix  to  two-terminal setups [Fig.~\ref{fig:fig1}], via Eq.~\eqref{eq:retardedgfleads} plugged into Eq.~\eqref{eq:rhokubo}, which is an unexplored route in studies arriving at {\em divergent conclusions}; and ({\em ii}) properly construct~\footnote{Note that many Keldysh density matrix-based studies of current-driven $\langle \hat{\mathbf{s}}_i \rangle$~\cite{Reynoso2006} or thereby induced spin torque~\cite{Haney2007,Kalitsov2017} in two-terminal Landauer 
		setup use only $\rho^\mathrm{surf}_\mathrm{Keldysh}$ [Eq.~\eqref{eq:rhokeldyshlrsurf}], while evading computation of $\rho^\mathrm{sea}_\mathrm{Keldysh}$ [Eq.~\eqref{eq:rhokeldyshlrsea}], which can produce  results {\em violating} the gauge invariance~\cite{Mahfouzi2013,Christen1996}.}   $\hat{\rho}^\mathrm{sea}_\mathrm{Keldysh}$ contribution to the Keldysh density matrix   [Eq.~\eqref{eq:rhokeldyshlrsea}], which requires using (a rarely computed~\cite{Belashchenko2019,Mahfouzi2013}) \mbox{$\hat{G}^r_{V_b} = \left[E  - {\hat{H}} - eU_i -\hat{\Sigma}_L^r -\hat{\Sigma}_R^r \right]^{-1}$} in Eq.~\eqref{eq:rhokeldyshlrsea} that includes 
	the voltage drop $e U_i$  across the CA region [Fig.~\ref{fig:fig1}]. Note that $\hat{G}^r_{V_b}$ is markedly different from $\hat{G}^r$ [Eq.~\eqref{eq:retardedgfleads}] used in all other terms of Kubo Eq.~\eqref{eq:rhokubo} or Keldysh Eq.~\eqref{eq:rhokeldyshlr}. Since a voltage drop {\em cannot}~\cite{Payne1989} be justified for clean CA region hosting ballistic charge transport, this also suggests that the computation of local transport quantities always requires the introduction of real-space disorder, even though disorder averaging is often avoided  due to the high computational cost of repeated integrations over the Fermi-sea~\cite{Belashchenko2019} (recently developed spectral algorithms could be used to mitigate this problem \cite{Castro2024}).  Bulk properties can still be  extracted~\cite{Wang2016a} from two-terminal systems with CA region of finite length by computing local quantities at some distance~\cite{Wang2016a,Belashchenko2023} away from the CA-region/lead interface (such as, at the shaded middle hexagons in Fig.~\ref{fig:fig1}).

	\begin{figure*}
		\begin{center}
			\includegraphics[width=0.96\linewidth]{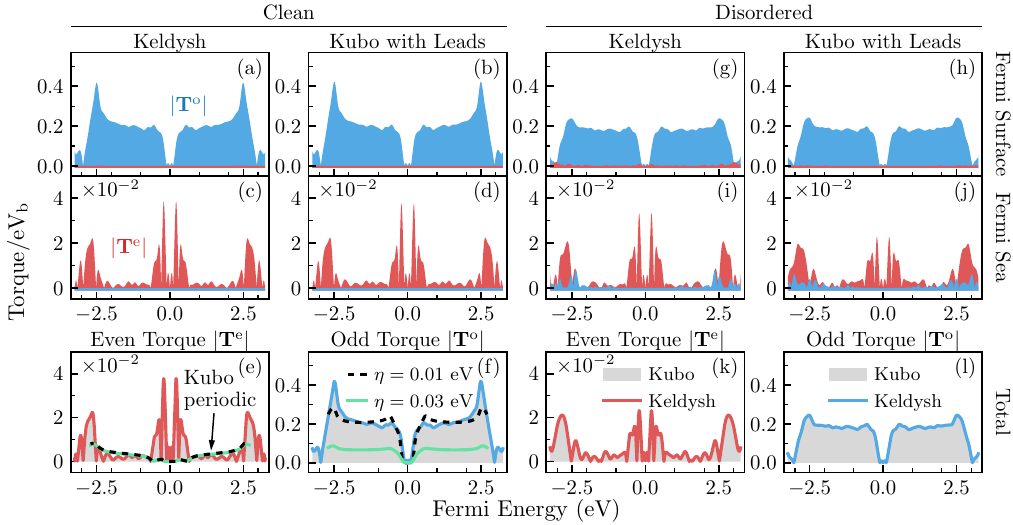}
		\end{center}
		\vspace{-0.3in}	\caption{Similar to Fig.~\ref{fig:fig2}, but for  even $\mathbf{T}^e$ (or damping-like~\cite{Ralph2008,Belashchenko2019}) and odd $\mathbf{T}^o$ (or field-like~\cite{Ralph2008,Belashchenko2019}) SO torques. Each row was obtained by tracing the torque operator~\cite{Belashchenko2019,Nikolic2018} with (a),(b),(g),(h) Fermi-surface; or (c),(d),(i),(j) Fermi-sea density matrix; as well as (e),(f),(k),(l) total density matrix within Kubo Eq.~\eqref{eq:rhokubo} vs. Keldysh Eq.~\eqref{eq:rhokeldyshlr}. In panels (a)--(f), the CA region   in Fig.~\ref{fig:fig1} is clean, while in panels (g)--(l) it contains Anderson disorder of strength \mbox{$D=0.5$ eV}. Panels (e) and (f) show additional (black and green) curves obtained from conventional~\cite{Freimuth2014,Mahfouzi2018,Mahfouzi2020,Ghosh2018,MedinaDuenas2024} Kubo calculations on periodic lattices [i.e., by using Eq.~\eqref{eq:retardedgf} plugged into  Eq.~\eqref{eq:rhokubo}]. The parameters in Eq.~\eqref{eq:hamiltonian} read as \mbox{$\lambda_\mathrm{RSO} = J_{\text{sd}} = 0.3$ eV}. Calculations with disorder employ $200$ configurations.}
		\label{fig:fig3}
	\end{figure*}

	To demonstrate such an equivalence, we compute the SH   current [Fig.~\ref{fig:fig2}] and SO torque [Fig.~\ref{fig:fig3}] densities via Kubo and Keldysh routes for the  same system---a graphene sheet with  both SO coupling and magnetic ordering [Fig.~\ref{fig:fig1}].  The effective TB Hamiltonian (whose parameters can be fitted to first-principles calculations~\cite{Zollner2020} or experimental data~\cite{Perkins_2024}) is given by
	\begin{eqnarray}\label{eq:hamiltonian}
		\hat{H}  & = &  
		-t \sum_{\langle i, j\rangle, \sigma} 
		\hat{c}^\dagger_{i\sigma} \hat{c}_{j\sigma} + \frac{2i\lambda_{\rm RSO}}{3}
		\sum_{\langle i,j \rangle,\sigma \neq \sigma'}
		{[\hat{\bm \sigma} \times 
			{\boldsymbol d}_{ij}]}_{\sigma \sigma'} 
		\hat{c}^\dagger_{i \sigma} \hat{c}_{j \sigma'}  \nonumber \\ 
		\mbox{} && + 
		\sum_{i,\sigma,\sigma^\prime} \hat{c}_{i\sigma}^{\dagger} \bigg( \varepsilon_{i}\delta_{\sigma\sigma^{\prime}}+J_{{\rm sd}}[\mathbf{m}_i \cdot\hat{\bm{\sigma}}]_{\sigma\sigma^{\prime}} \bigg) \hat{c}_{i\sigma^{\prime}}
		.
	\end{eqnarray}
	Physically, such spin-dependent interactions can be introduced in graphene via proximity effects from an overlayer of 2D ferromagnetic insulator and underlayer of 2D semiconductor material~\cite{Zollner2020}. 
	Here $\hat{c}^\dagger_{i\sigma}$ ($\hat{c}_{i\sigma}$) creates (annihilates) an electron at a site $i$ with spin $\sigma=\uparrow,\downarrow$; $\hat{\bm \sigma} = (\hat{\sigma}_x,\hat{\sigma}_y,\hat{\sigma}_z)$  is the vector of the Pauli matrices; \mbox{$t=2.7$ eV} is the nearest-neighbor (NN) hopping; the sum $\langle i,j \rangle$ goes over all pairs of  NN sites; $\varepsilon_i \in [-D/2,D/2]$ is a uniform random variable introducing Anderson disorder on each site;  $J_{\rm sd}$ is $sd$ exchange coupling between conduction electrons and  proximity-induced localized magnetic moments  described by a classical unit  vector  $\mathbf{m}_i$. While other SO   effects can be induced in graphene ~\cite{Zollner2020,2DSpintronics_CMP_Encyclopedia},  we focus on symmetry-breaking Rashba SO coupling of strength $\lambda_\mathrm{RSO}$, where  $\mathbf{d}_{ij}$ is the unit vector along the direction connecting NN sites $i$ and $j$. The $k$-point sampling along the transverse direction is implemented~\cite{Liu2012d} with   hoppings connecting sites along the lower and upper edge of armchair nanoribbon  multiplied by the phase $e^{ik_{y}W}$. All results  in Figs.~\ref{fig:fig1}(c), ~\ref{fig:fig2} and ~\ref{fig:fig3} are averaged~\cite{Liu2012d,Thygesen2005} over the transverse wavevector  using $\langle \hat{O} \rangle 
	= \frac{W}{2\pi} \int_{-\pi/W}^{\pi/W} \! dk_y\, \langle \hat{O} \rangle_{k_y}$. Though not essential for the Kubo vs. Keldysh equivalence, the $k$-sampling makes the system behave as infinite along the $y$-axis and thus yields bulk-like~\cite{Liu2012d} behavior  [see its density of states in Fig.~S1(d) of the SM~\footnotemark[1]].

	In order to reconcile the Kubo and Keldysh approaches, we start by recalling that  for pure Fermi-surface transport properties their equivalence is well-established from long ago~\cite{Baranger1989}. This has been amply confirmed~\footnote{Nevertheless, confusion occasionally arises 
		in the literature, such as in Fig.~6 of Ref.~\cite{Fan2021} where the Kubo conductivity incorrectly does not match quantized conductance from Keldysh approach} for, e.g.,  conductance~\cite{Baranger1989,Fisher1981,Kane1988} of two-terminal systems~\cite{Nikolic2001b}, including graphene~\cite{Castro2023,Castro2024} that we also revisit in the SM~\footnotemark[1].  
	Since the conductance formulas~\footnotemark[1] 
	for  \mbox{$G_\mathrm{Kubo} \equiv G_\mathrm{Keldysh}$} are essentially the expectation value of the {\em total} current operator~\cite{Baranger1989,Mahfouzi2013} in the right lead divided by the voltage drop (i.e.,  \mbox{$G=\langle \hat{I}_R \rangle/V_b$},  where \mbox{$\langle \hat{I}_R \rangle = \mathrm{Tr}\, [\hat{\rho}^\mathrm{surf}_\Box \hat{I}_R]$} and $\Box$=Kubo or $\Box$=Keldysh), this implies equivalence~\cite{Baranger1989} of   $\textrm{Tr}[\hat{I}_{R}\hat{\rho}_{\mathrm{Kubo}}^{\mathrm{surf}}]$ and   $\textrm{Tr}[\hat{I}_{R}\hat{\rho}_{\mathrm{Keldysh}}^{\mathrm{surf}}]$  on the proviso  that the retarded GF in Eq.~\eqref{eq:retardedgfleads} is used. %NOTE: there is no need to say that the equations have to be applied to the same system, that goes without saying! 
	This suggests that the {\em divergent conclusions} in recent studies of spin-dependent transport stem from  attempts to calculate  expectation values of {\em local} quantities that require  additional traces of their operators with the \textit{Fermi-sea terms}, $\hat{\rho}^\mathrm{sea}_\mathrm{Kubo}$ [Eq.~\eqref{eq:rhokubosea}] or $\hat{\rho}^\mathrm{sea}_\mathrm{Keldysh}$ [Eq.~\eqref{eq:rhokeldyshlrsea}]. To investigate this issue, in Fig.~\ref{fig:fig2} we first consider the expectation value of spin (Hall) current density $\langle \hat{j}^{S_z}_y \rangle$, with the corresponding operator being~\cite{Wang2016a} $\hat{j}^{S_z}_y =\frac{e}{2}(\hat{v}_y \hat{\sigma}_z +  \hat{\sigma}_z \hat{v}_y)$. In order to capture the bulk behavior,
	%to assist numerical convergence (the equivalence we demonstrate is of course more general), 
	this local transport quantity is computed and 
	averaged over a small area in the middle of the CA region as denoted  (yellow hexagons) in Fig.~\ref{fig:fig1}. Note that  $\langle \hat{j}^{S_z}_y \rangle$, or the SH conductivity $\sigma_\mathrm{SH}=\langle \hat{j}^{S_z}_y \rangle/E_x$, in the pure Rashba model (i.e., with $J_{\text{sd}}=0$) treated by the standard Kubo calculations  with periodic boundary conditions has zero contribution from the Fermi-surface due to vertex corrections when both Rashba-split bands are occupied \cite{Dimitrova2005,Milletari2017}. Interestingly, our calculations for $J_{\text{sd}}\neq 0$ [gray curve in Fig.~\ref{fig:fig2}] produce a relatively small nonzero value, which we attribute to artifacts of disorder averaging. The suppression of the Fermi-surface contribution to $\hat{j}^{S_z}_y$ in our results is  nontrivial since the interplay of Rashba SOC and exchange coupling is expected to generate robust extrinsic SHE via skew scattering, as shown by Boltzmann and Kubo  calculations~\cite{Offidani_2018} (enhanced intervalley scattering due to the nature of our short-range disorder landscape~\cite{Offidani_2018}, as well as the non-diffusive nature of our transport simulations, are likely explanations for this behavior).  By contrast, the Keldysh calculations [Fig.~\ref{fig:fig2}(a),(b)] contain significant contributions from both the Fermi-surface and Fermi-sea. Although this suggests that $\hat{\rho}^\mathrm{surf}_\mathrm{Kubo} \neq \hat{\rho}^\mathrm{surf}_\mathrm{Keldysh}$ and $\hat{\rho}^\mathrm{sea}_\mathrm{Kubo} \neq \hat{\rho}^\mathrm{sea}_\mathrm{Keldysh}$, it is the sum of both contributions which carries the physical meaning. Indeed, we see that $\textrm{Tr}[\hat{\rho}_{\mathrm{\text{Keldysh}}}^{\mathrm{neq}} \hat{j}_{y}^{S_{z}}] \equiv  \textrm{Tr}[\hat{\rho}_{\mathrm{\text{Kubo}}}^{\mathrm{neq}} \hat{j}_{y}^{S_{z}}]$   perfectly match in Fig.~\ref{fig:fig2}(c). 
	
	Another local transport quantity that has been a source of {\em divergent conclusions} is the SO torque~\cite{Manchon2019}---an intensely studied phenomenon over the past decade due to its experimental~\cite{Kurebayashi2014} and technological relevance~\cite{Locatelli2014,Borders2017}.  In general, spin torques arise~\cite{Ralph2008,Petrovic2021a} due to exchange of spin angular momentum between flowing electrons and localized magnetization. Specifically, 
	injected unpolarized charge current, together with SO coupling, produces nonequilibrium spin density, 
	\mbox{$\langle \hat{\mathbf{s}}_i \rangle = \mathrm{Tr}\, [\hat{\rho}^\mathrm{neq}_\Box \hat{\bm \sigma}]$} where \mbox{$\hat{\mathbf{s}}_i = \hat{c}^\dagger \hat{\bm \sigma} \hat{c}_i$} is the spin operator, whose computation makes it possible to obtain local SO torque as \mbox{$\mathbf{T}_i = J_\mathrm{sd} \langle \hat{\mathbf{s}}_i \rangle \times \mathbf{m}_i$}. Both $\Box$=Kubo~\cite{Ado2017,Freimuth2014,Mahfouzi2018,Mahfouzi2020,MedinaDuenas2024,Sousa2020,Kurebayashi2014,Li2015,Lee2015,Ghosh2018} and $\Box$=Keldysh~\cite{Belashchenko2019,Dolui2020,Zollner2020,Kalitsov2017,Mahfouzi2016,Belashchenko2020,Nikolic2018}  density matrices have been frequently  used in SO torque calculations.  Similar to Fig.~\ref{fig:fig2}, we average local SO torque  over the
	middle of the CA region  indicated in Fig. 1 to obtain \mbox{$\mathbf{T}= \frac{1}{N}\sum_{i=1}^N \mathbf{T}_i = \mathbf{T}^e + \mathbf{T}^o$}.   Figure~\ref{fig:fig3}(a)--(f) shows the energy dependence of the even and odd components of $\mathbf{T}$, with magnetization fixed out of plane  \mbox{$\mathbf{M}=\sum_i \mathbf{m}_i \parallel \hat z$}, calculated for the clean system. Since both  formulations in Fig.~\ref{fig:fig3}(a)--(f) yield identical $\mathbf{T}^e_\mathrm{Kubo} \equiv \mathbf{T}^e_\mathrm{Keldysh}$ and $\mathbf{T}^o_\mathrm{Kubo} \equiv \mathbf{T}^o_\mathrm{Keldysh}$, this demonstrates how the Keldysh formula
	reproduces Fermi-sea-governed even (or damping-like~\cite{Ralph2008,Belashchenko2019}) SO torque $\mathbf{T}^e$ in the clean limit, that was previously considered arising only via the Kubo route~\cite{Kurebayashi2014,Li2015,Lee2015,Ghosh2018}. In Fig.~\ref{fig:fig3}(e),(f), we additionally show the even and odd SO torque produced by conventional usage~\cite{Freimuth2014,Mahfouzi2018,Ghosh2018,MedinaDuenas2024} of the Kubo Eq.~\eqref{eq:rhokubo} on periodic lattices. As expected, the particular choice of the {\em ad hoc} broadening  $\eta$ affects the results for $\mathbf{T}^o$ [Fig.~\ref{fig:fig3}(f)], diverging with $\eta \rightarrow 0$ which is unphysical~\cite{Kalitsov2017,Wu2010c}. Moreover, $\mathbf{T}^e$ which  is 
	independent of $\eta$ (and, therefore, considered ``intrinsic''~ \cite{Kurebayashi2014,Li2015,Lee2015}),   {\em deviates} substantially  from our Kubo$\equiv$Keldysh results for $\mathbf{T}^e$ on two-terminal systems [Fig.~\ref{fig:fig3}(e)]. Thus, the implementation of the Kubo(-Bastin)   Eq.~\eqref{eq:rhokubo} on two-terminal geometries that we develop here evades ambiguities due to choice of $\eta$ in finite-size periodic lattice calculations because two-terminal Landauer setups are infinite systems  with a {\em continuous}  energy spectrum~\cite{Waintal2024}  (as demanded also in the original derivations of Kubo~\cite{Kubo1957,Wu2010c,Kundu2009,Kamiya2013}) which ensures that dissipation is effectively introduced ~\cite{Giuliani2008,Petrovic2018,VarelaManjarres2023}.
	Our implementation also  mimics closely  experimental setups where a nonequilibrium state is introduced~\cite{Wu2010c,Baranger1989,Kundu2009,Kamiya2013} by injecting current through the leads or by applying a voltage difference $V_b$ between them, rather than by applying electric field. In the disordered case, Fig.~\ref{fig:fig3}(g) shows that $\hat{\rho}_\mathrm{Keldysh}^\mathrm{surf}$  produces additional contributions to damping-like $\mathbf{T}^e$, which we attribute to the often overlooked skew-scattering-induced damping-like SO torque~\cite{Sousa2020,Zollner2020,Veneri_22,Zhang2023a}. Also, $\hat{\rho}_\mathrm{Keldysh}^\mathrm{sea}$ and $\hat{\rho}_\mathrm{Kubo}^\mathrm{sea}$ produce [Fig.~\ref{fig:fig3}(i),(j)] additional contributions to $\mathbf{T}^o$. Despite these  differences in specific 
	contributions, their sums produce identical $\textrm{Tr}[\hat{\rho}_{\mathrm{\text{Keldysh}}}^{\mathrm{neq}} \hat{T}_i^{e,o}] \equiv  \textrm{Tr}[\hat{\rho}_{\mathrm{\text{Kubo}}}^{\mathrm{neq}} \hat{T}_i^{e,o}]$  in Fig.~\ref{fig:fig3}(k),(l). This, together with the results of  Fig.~\ref{fig:fig2}(c),  completes our proof of equivalence. 
	
	The numerical frameworks developed and validated   here demonstrate an unambiguous route to study generic spin-charge transport phenomena in the linear-response regime of realistic systems, in addition to    resolving a debate in spintronics over the proper usage of Kubo and Keldsyh formulas.    Our findings also suggests that assigning a unique and special  physical meaning~\cite{Freimuth2014,Kurebayashi2014,Li2015,Lee2015} to the Fermi-sea term in Kubo [Eq.~\eqref{eq:rhokubo}]  requires careful scrutiny, as the decomposition of  density matrix into Fermi-surface and Fermi-sea contributions is not unique~\footnotemark[1] and, as seen through the demonstrated Kubo vs. Keldysh equivalence, there are many possible~\footnotemark[1] (and rather mundane looking) forms of the Fermi-sea term within the Keldysh formalism [Eq.~\eqref{eq:rhokeldyshlr}]. The  particular expression to be used in practical calculations is a matter of computational convenience~\footnote{Note that in the ``symmetrized  decomposition'' of Ref.~\cite{Bonbien2020}, the integrand in Eq.~\eqref{eq:rhokubosea} is not analytic in the upper complex plane, which introduces additional computational complexity. In some other decompositions, and for the integration in Eq.~\eqref{eq:rhokeldyshlrsea}, many efficient complex contour-based algorithms have been developed~\cite{Brandbyge2002,Ozaki2007,Areshkin2010}, see the SM for more details.}. 
	
	B.~K.~N.  was supported by the US NSF through the  University of Delaware Materials Research Science and Engineering Center, DMR-2011824. The supercomputing time was provided by DARWIN (Delaware Advanced Research Workforce and Innovation Network), which is supported by NSF Grant No. MRI-1919839. A.~F. acknowledges the partial support from a Royal Society University Research Fellowship.
	
%	\bibliography{references}

%

\end{document}